Pressure study of the new iron-based superconductor $K_{0.8}Fe_2Se_2$

Y. Kawasaki [a, b, c], Y. Mizuguchi [a, c], K. Deguchi [a, b, c], T. Watanabe [a, b], T. Ozaki [a, c],

S. Tsuda [a, c], T. Yamaguchi [a, c], H. Takeya [a, c], Y. Takano [a, b, c]

[a] National Institute for Materials Science, 1-2-1 Sengen, Tsukuba 305-0047, Japan

[b] University of Tsukuba, 1-1-1 Tennoudai, Tsukuba, Ibaraki 305-0006, Japan

[c] JST-TRIP, 1-2-1 Sengen, Tsukuba 305-0047, Japan

In 2008, Hsu *et al.* reported superconductivity at 8 K in anti-PbO-type FeSe [1]. Because of the simple crystal structure, FeSe has attracted the attention of researchers. Furthermore, the $T_c$ of FeSe dramatically increases with applying an external pressure and reaches 37 K at 4 GPa [2 - 5]. The enhancement of $T_c$ under pressure can be explained with the change of local crystal structure such as anion height [6]. Recently, new iron-based superconductor $A_xFe_2Se_2$ (A: K, Rb, Cs, Tl) was found to exhibit superconductivity around 30 K [7 - 13]. Intercalating cation into the inter-layer site would tune the local crystal structure and achieves high-$T_c$ superconductivities in the FeSe layers. To investigate a possibility of higher $T_c$ in this system, we performed pressure experiment of $K_{0.8}Fe_2Se_2$.

Single crystals of $K_{0.8}Fe_2Se_2$ were prepared by melting FeSe precursor and K as described in Ref. 9. Resistivity measurements were performed using the four-terminal method from 300 to 2 K. Fabrication of four terminals was carried out in a globe bag filled with argon gas to avoid a degradation of the sample; gold wires of 25 $\mu$m in diameter were attached to the sample using silver paste. We used the BeCu/NiCrAl hybrid-type clamped piston-cylinder cell to generate the hydrostatic pressures. The sample was immersed in a fluid pressure transmitting medium of Fluorinert (FC70:FC77 = 1:1) and sealed in a Teflon cell. The precise pressure at low temperature was estimated from the superconducting transition temperature of Pb.

Figure 1 shows the temperature dependence of resistivity from 300 to 2 K for $K_{0.8}Fe_2Se_2$ under high pressure. The resistivity-temperature curve shows a hump around 200 K as at ambient pressure. With increasing pressure, resistivity decreased and the hump was suppressed. But the temperature at which the hump appeared did not change remarkablely under pressure. Figure 2 is an enlargement of superconducting transition. The $T_c^{onset}$ was determined to be the temperature where the resistivity deviates from the linear extrapolation line as shown in Fig. 2. A sharp transition was observed at 0.85 GPa and the $T_c^{onset}$ and $T_c^{zero}$ were estimated to be 33 K and 31.5 K, respectively. These values are almost the same as $T_c$ under ambient pressure. The $T_c^{onset}$ increases with increasing pressure and reaches 36.6 K at 2.03 GPa. Although $T_c^{onset}$ increases with increasing pressure, the $T_c^{zero}$ decreases. The estimated $T_c^{onset}$ and $T_c^{zero}$ are plotted in Fig. 3 as a function of pressure. Recently, J. Guo *et al*. reported that both the $T_c^{onset}$ and $T_c^{zero}$ of $K_{0.8}Fe_{1.7}Se_2$ decreased with increasing pressure and superconductivity disappeared under pressure above 9.2 GPa [13]. In our study, the $T_c^{zero}$ was suppressed with increasing pressure as well as in Ref. 13, but the $T_c^{onset}$ was clearly enhanced. Furthermore, the $T_c^{onset}$ seems to exhibit a further enhancement of $T_c$ under higher pressure. The $T_c$ under pressure could be correlated with local crystal structure as well as in FeSe. For a greater understanding of the superconducting properties of $K_{0.8}Fe_2Se_2$ under pressure, structural analysis under pressure is required. Furthermore, it is also important to investigate the pressure effect of the other $A_xFe_2Se_2$ superconductors.

In conclusion, we investigated the pressure effect of $K_{0.8}Fe_2Se_2$ single crystal. The $T_c^{onset}$, 33 K at ambient pressure, was enhanced with applying an external pressure and reached 36.6 K at 2.03 GPa. In contrast, $T_c^{zero}$ decreased with increasing pressure. To understand an intrinsic property of $K_{0.8}Fe_2Se_2$ under pressure, preparation of crystals stable in air and structural analysis under pressure are needed.

**Acknowledgement**

This work was partially supported by Grant-in-Aid for Scientific Research (KAKENHI).

Figure captions

Fig. 1. Temperature dependence of resistivity for $K_{0.8}Fe_2Se_2$ under pressure.

Fig. 2. Enlargement of superconducting transition for $K_{0.8}Fe_2Se_2$ under pressure.

Fig. 3. Pressure dependence of superconducting transition temperature for $K_{0.8}Fe_2Se_2$. Open circles are the data at ambient pressure [9].

Fig. 1

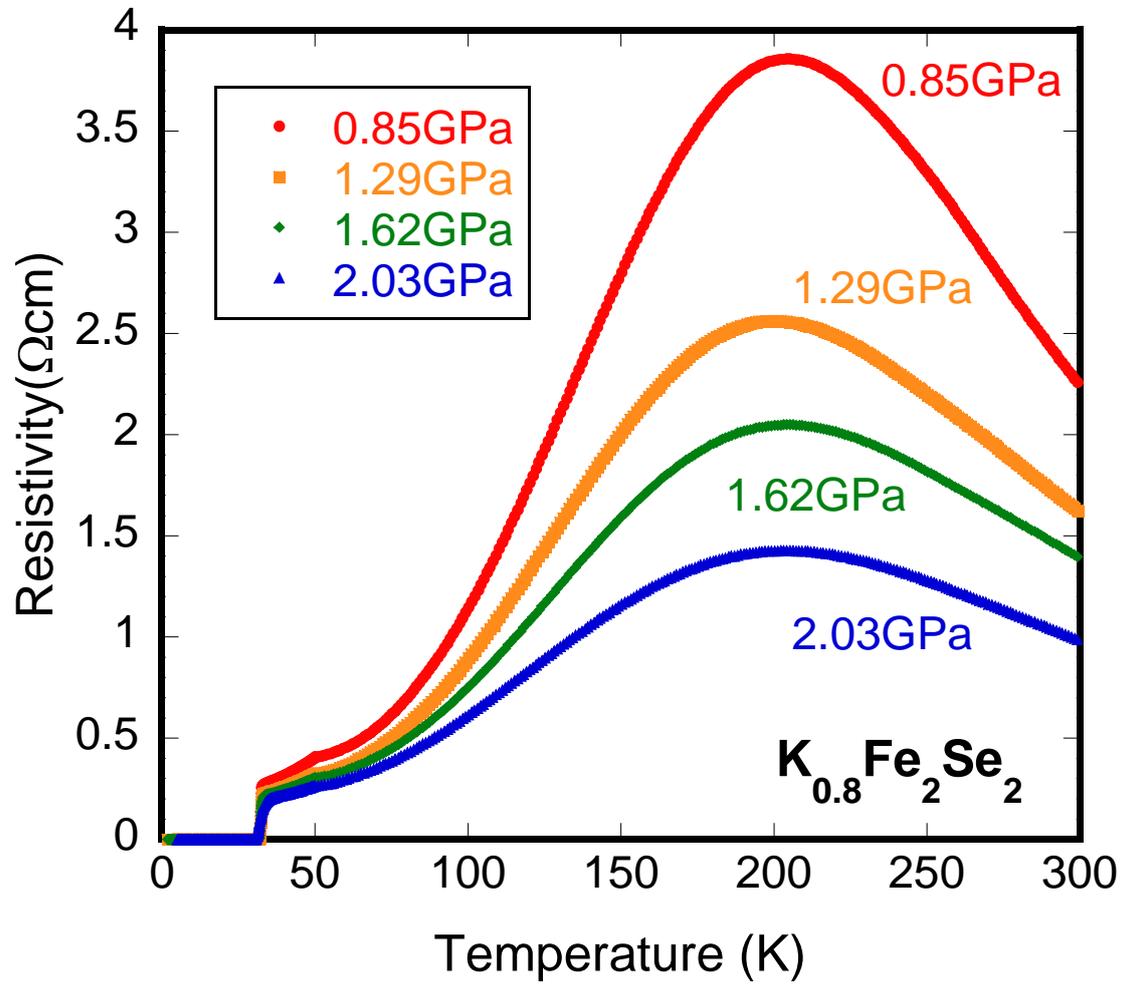

Fig. 2.

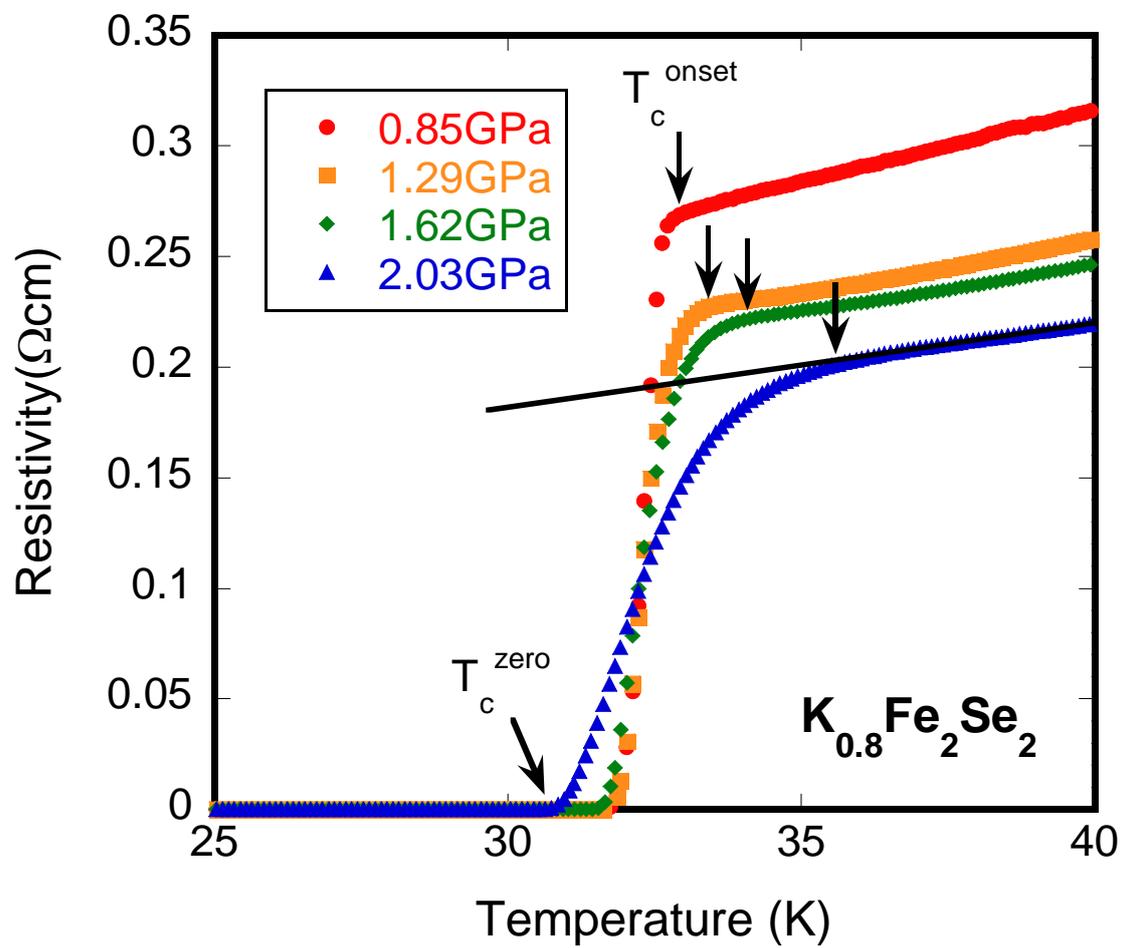

Fig. 3.

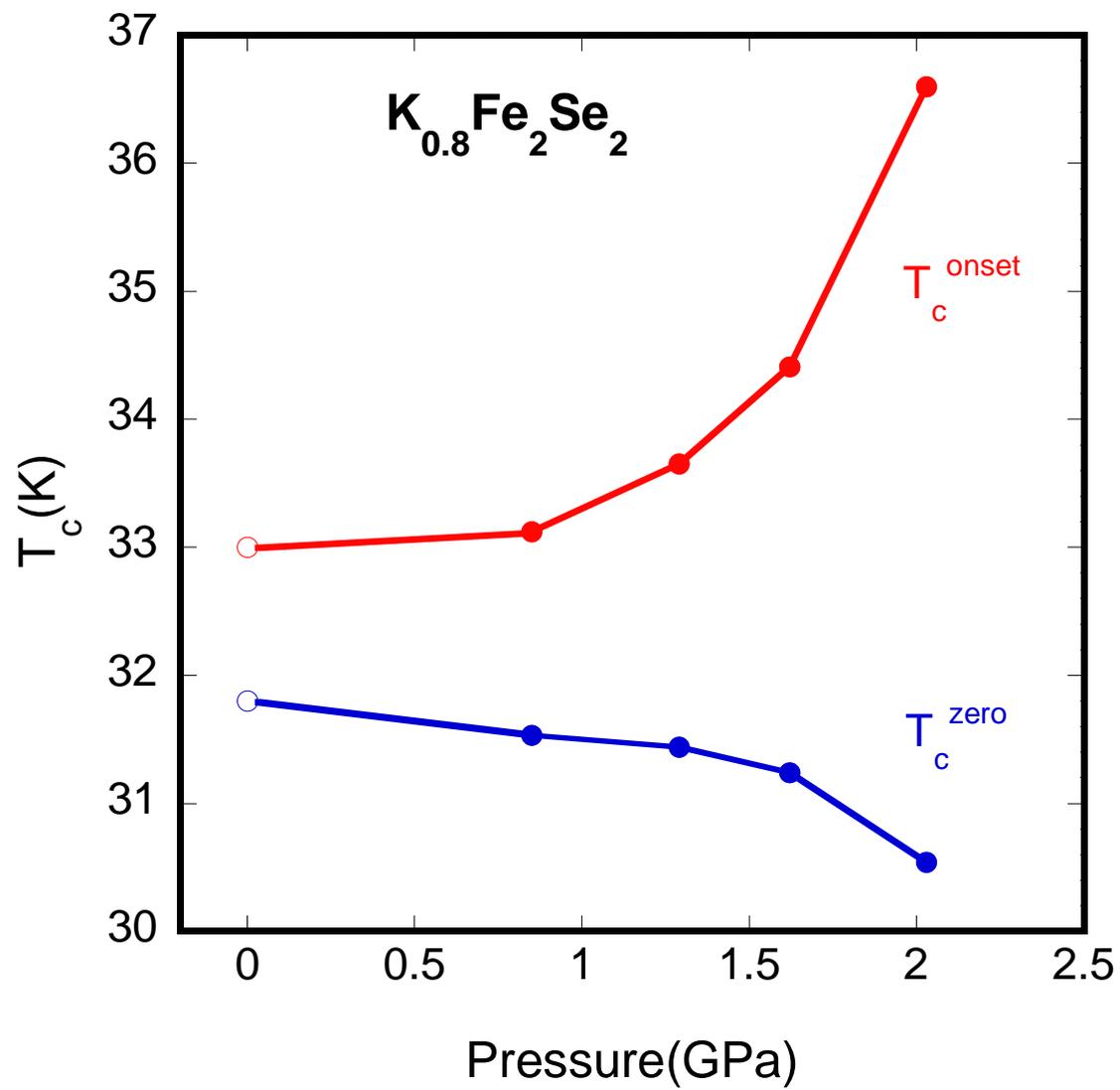